\documentclass[pre,aps,twocolumn,,amsmath,amssymb,showpacs,showkeys]{revtex4}

\usepackage{graphicx}
\usepackage{color}
\definecolor{r1}{rgb}{0.0,0.0,0.0}
\newcommand{\hh}{{\mathcal{H}}}
\newcommand{\lnp}{{\mathcal{L}}}
\newcommand{\lsa}{{\mathcal{L}}_{s.a.}}
\newcommand{\lsp}{{\mathcal{L}}_{+}}

\newcommand{\pen}{\openone}
\newcommand{\Tr}{{\mathrm{Tr}}}
\newcommand{\id}{{\mathrm{id}}}

\newcommand{\htk}{{\hat{K}}}
\newcommand{\hta}{{\hat{A}}}
\newcommand{\htb}{{\hat{B}}}
\newcommand{\htc}{{\hat{C}}}
\newcommand{\htg}{{\hat{G}}}
\newcommand{\htx}{{\hat{X}}}
\newcommand{\hth}{{\hat{H}}}
\newcommand{\htd}{{\hat{D}}}
\newcommand{\htu}{{\hat{U}}}
\newcommand{\htr}{{\hat{\rho}}}
\newcommand{\htl}{\hat{\lambda}}
\newcommand{\hto}{\hat{\omega}}
\newcommand{\htvr}{\hat{\varrho}}
\newcommand{\htom}{\hat{\Omega}}
\newcommand{\htvp}{\hat{\eta}}
\newcommand{\bsg}{\boldsymbol{\sigma}}
\newcommand{\blm}{\boldsymbol{\lambda}}
\newcommand{\hsg}{\hat{\sigma}}
\newcommand{\blj}{\mathbf{J}}

\newcommand{\veps}{\varepsilon}

\newcommand{\bau}{\boldsymbol{\tau}}
\newcommand{\mib}{{\mathbf{B}}}

\begin{document}
\clearpage
\preprint{}

\title{Jarzynski equality for quantum stochastic maps}
\author{Alexey E. Rastegin$^{1}$ and Karol \.{Z}yczkowski$^{2,3}$}

\affiliation{
\mbox{$^1$Dept. of Theoretical Physics, Irkutsk State University, Gagarin Bv. 20,
Irkutsk 664003, Russia}\\
\mbox{$^2$Institute of Physics, Jagiellonian University, ul.\ Reymonta 4, 30-059 Krak\'ow, Poland}\\
\mbox{$^3$Center for Theoretical Physics, Polish Academy of Sciences, al.\ Lotnik\'ow 32/46, 02-668 Warszawa, Poland}\\
}

\date{Januar 23, 2014}

\begin{abstract}

Jarzynski equality and related fluctuation theorems can be
formulated for various setups. Such an equality was recently
derived for nonunitary quantum evolutions described by unital
quantum operations, i.e., for completely positive, trace-preserving 
maps, which preserve the maximally mixed state. We
analyze here a more general case of arbitrary quantum operations
on finite systems and derive the corresponding form of the
Jarzynski equality. It contains a correction term due to
nonunitality of the quantum map. Bounds for the relative size of
this correction term are established and they are applied for
exemplary systems subjected to quantum channels acting on a
finite-dimensional Hilbert space.

\end{abstract}
\maketitle

\pagenumbering{arabic}
\setcounter{page}{1}

\section{Introduction}\label{sec1}

Recent theoretical and experimental advances in dealing with small
quantum systems has led to a growing interest in their mechanics and
thermodynamics \cite{jareq11}. A certain amount of progress has been
connected with studies of the Jarzynski equality \cite{jareq97a}
and related fluctuation theorems
\cite{talhag07,talhag09,cht11,vvliet12}. Recent attention is
mainly focused on the quantum version  of these results. Quantum
analogues of the Jarzynski equality were first studied by
Kurchan \cite{kurchan00} and Tasaki \cite{tasaki99}. Since then
various topics connected with the fluctuation relations and the
range of their validity and applicability were investigated.

There exist many ways to approach the Jarzynski equality
\cite{jareq97b,crooks99,jareq00,seif04,scholl06,TM07,Cr08,cohen12,Pa12}.
Most of them are based on a dynamical description within an
infinitesimal time scale. Making use of the perturbation approach,
the author of Ref. \cite{vvliet12} analyzed quantum fluctuation
and work-energy theorems that focus on the time-reversal symmetry.
We will advocate here a different approach applicable for systems
which can be described by discrete quantum operations.

The formalism of quantum operations is one of the basic tools in
studying dynamics of open quantum systems
\cite{nielsen,bengtsson}. Fluctuation theorems for open quantum
systems were recently considered in Refs.
\cite{vedral12,kafri12,albash,rast13,deffner13}. In particular,
some results have been shown to be valid in the case of unital
quantum operations, while the general case of quantum systems with
time evolution described by nonunital stochastic maps remained
not fully understood.

The main goal of this study is to relax the assumption of
unitality and to generalize previous results for the entire class
of stochastic maps, also called quantum channels. Another task of
the work is to introduce a model discrete quantum dynamics acting
on a $N$-dimensional system, which forms a useful generalization
of the amplitude damping channel acting on a two-level system.
This nonunital map channel and its extensions describe effects of
energy loss in quantum systems due to an interaction with an
environment \cite{nielsen,bengtsson}. Investigation of possible
effects due to deviations from unitality of the map become
relevant in the context of possible experimental tests of quantum
fluctuation theorems.

Experimental study of fluctuation relations is easier in the
classical regime \cite{cht11}. Original formulations of the
Jarzynski equality and the Crooks theorem were tested in
experiments \cite{liph02,busta05,dcpr05,npl10,sytmap12}. On the
other hand, experimental investigation of quantum fluctuation
relations is still forthcoming, although some possible
experimental schemes were already discussed
\cite{hskde08,dchfgv13,mdcp13,cbkzh13}. Existing proposals often
deal with a single particle undergoing an unitary time evolution.
Furthermore, current efforts to construct devices able to process
quantum information might offer new possibilities to test quantum
fluctuation relations. Notably, quantum systems are very sensitive
to interaction with an environment. In this regard, fluctuations
in systems with an arbitrary form of quantum evolution deserve
theoretical analysis. Therefore, we do not focus our attention on
a specific class of unital channels, but we study the most general
form of arbitrary quantum operations.

The original formulations of the Jarzynski equality and the
Tasaki--Crooks fluctuation theorem remain valid under the
assumption that changes of the system state are represented by a
unital quantum operation \cite{albash,rast13}. Attention to
bistochastic maps is natural, when we deal with the Tasaki--Crooks
fluctuation theorem. Indeed, its formulation involves both the
forward quantum channel and its adjoint. If the latter channel
preserves the trace, then the former one is necessarily unital.

Meantime, nonunital quantum channels are of interest in various
respects. In this work we provide a formulation of the Jarzynski
equality for arbitrary quantum operations. The contribution of our
paper is twofold. First, we formulate a generalization of the
Jarzynski equality for a nonunital quantum channel. Second, we
investigate the problem for which the
standard Jarzynski equality remains valid nonunital quantum channels.

This paper is organized as follows. In Section \ref{sec2}, we
introduce basic definitions and recall relevant results. The
special case of unital channels and bistochastic maps is analyzed
in Sec. \ref{sec3}. In Sec. \ref{sec4}, we characterize
nonunitality of an arbitrary quantum stochastic map while in Sec.
\ref{sec5} we generalize the corresponding Jarzynski equality for
this class of maps and derive Eq. (\ref{jareq0g}) --- a key result
of the paper. Several examples of nonunital quantum channels
acting on two and three-level system are analyzed in Sec.
\ref{sec6}. We investigate also a general model of nonunitary
dynamics described in an arbitrary finite-dimensional Hilbert
space which can be considered as a generalization of the amplitude
damping channel.

\section{Definitions and notation}\label{sec2}

Let $\lnp(\hh)$ denote the space of linear operators on
$N$-dimensional Hilbert space $\hh$. By $\lsa(\hh)$ and
$\lsp(\hh)$, we respectively mean the real space of Hermitian
operators and the set of positive ones. For arbitrary
$\hta,\htb\in\lnp(\hh)$, we define their Hilbert--Schmidt inner
product by \cite{watrous1}
\begin{equation}
\langle\hta{\,},\htb\rangle_{\mathrm{hs}}
:=\Tr(\hta^{\dagger}\htb)
\ . \label{hsdef}
\end{equation}
This product induces the norm
$\|\hta\|_{2}=\langle\hta{\,},\hta\rangle_{\mathrm{hs}}^{1/2}$.
For any $\hta\in\lnp(\hh)$, we put $|\hta|\in\lsp(\hh)$ as a
unique positive square root of $\hta^{\dagger}\hta$. The
eigenvalues of $|\hta|$ counted with multiplicities are the
singular values  of $\hta$, written $s_{j}(\hta)$. For all real
$p\geq1$, the Schatten $p$ norm is defined as \cite{watrous1}
\begin{equation}
\|\hta\|_{p}:=\Bigl(\Tr\bigl(|\hta|^{p}\bigr)\Bigr)^{1/p}=
\Bigl(\sum\nolimits_{j=1}^{N} s_{j}(\hta)^{p}{\,}\Bigr)^{1/p}
\ . \label{schnd}
\end{equation}
This family includes the trace norm $\|\hta\|_{1}=\Tr|\hta|$ for
$p=1$, the Hilbert--Schmidt (or Frobenius) norm
$\|\hta\|_{2}=\bigl(\Tr(\hta^{\dagger}\hta)\bigr)^{1/2}$ for
$p=2$, and the spectral norm
$\|\hta\|_{\infty}=\max\bigl\{s_{j}(\hta):{\>}1\leq{j}\leq{N}\bigr\}$
for $p=\infty$. For all $q>p\geq1$, we have
\begin{equation}
\|\hta\|_{q}\leq\|\hta\|_{p}
\ . \label{npqr}
\end{equation}
This relation is actually a consequence of theorem 19 of the
classical book of Hardy, Littlewood, and Polya \cite{hardy}.

For any state of the $N$-level system we are going to use the
Bloch-vector representation, as it might be linked to experimental
data \cite{kgka05}. By $\htl_{j}\in\lsa(\hh)$,
$j=1,2,\ldots,N^{2}-1$, we denote the generators of
${\mathrm{SU}}(N)$ which satisfy $\Tr(\htl_{j})=0$ and
\begin{equation}
{\Tr}{\bigl(\htl_{i}\htl_{j}\bigr)}=2\delta_{ij}
\ . \label{lij2}
\end{equation}
The factor $2$ in Eq. (\ref{lij2}) is rather traditional and may
be chosen differently. Each traceless operator $\htx\in\lsa(\hh)$
can be then represented in terms of its Bloch vector as
\cite{bengtsson,kgka05}
\begin{equation}
\htx=\frac{1}{2}{\,}\sum\nolimits_{j=1}^{N^{2}-1}\tau_{j}\htl_{j}
\ , \qquad
\tau_{j}={\Tr}{\bigl(\htx\htl_{j}\bigr)}
\ . \label{blrx}
\end{equation}
Thus, we represent a traceless Hermitian $\htx$ by means of the
corresponding $(N^{2}-1)$-dimensional real vector
$\bau=\bigl(\tau_{1},\tau_{2},\ldots,\tau_{N^{2}-1}\bigr)$. For
the case $N=2$, the generators are the standard Pauli matrices
$\hsg_{j}$, where $j=1,2,3$. In the case $N=3$, the eight
Gell-Mann matrices are commonly used. In $N$-dimensional space
$\hh$, the completely mixed state is expressed as
\begin{equation}
\htr_{*}=\frac{1}{N}{\,}\pen \ ,
 \label{cmab}
\end{equation}
where $\pen$ is the identity operator on $\hh$. For a given
density matrix $\htr$, the operator $\htr-\htr_{*}$ is traceless,
whence a Bloch representation of $\htr$ follows Refs.
\cite{bengtsson,kgka05}.

Let us consider a linear map
$\Phi:{\>}\lnp(\hh_{A})\rightarrow\lnp(\hh_{B})$ that takes
elements of $\lnp(\hh_{A})$ to elements of $\lnp(\hh_{B})$. This
map is called positive if $\Phi(\hta)\in\lsp(\hh_{B})$ whenever
$\hta\in\lsp(\hh_{A})$ \cite{bhatia07}. To describe physical
processes, linear maps have to be completely positive
\cite{nielsen,bengtsson}. Let $\id_{R}$ be the identity map on
$\mathcal{L}(\hh_{R})$, where the space $\hh_{R}$ is assigned to a
reference system. The complete positivity implies that the map
$\Phi\otimes\id_{R}$ is positive for any dimension of the
auxiliary space $\hh_{R}$. The authors of Ref. \cite{frth04}
examined an important question, whether the dynamics of open
quantum systems is always linear. Further, we will consider only
completely positive linear maps. A completely positive map $\Phi$
can be written by an operator-sum representation,
\begin{equation}
\Phi(\hta)=\sum\nolimits_{n} \htk_{n}\hta\htk_{n}^{\dagger}
\ . \label{opsm}
\end{equation}
Here, the Kraus operators $\htk_{n}$ map the input space $\hh_{A}$
to the output space $\hh_{B}$. When physical process is
closed 
and the probability is conserved, the map preserves the trace,
$\Tr\bigl(\Phi(\hta)\bigr)=\Tr(\hta)$. This relation satisfied for
all $\hta\in\lnp(\hh_{A})$ is equivalent to the following
constraint for the set of the Kraus operators:
\begin{equation}
\sum\nolimits_{n} \htk_{n}^{\dagger}\htk_{n}=\pen_{A}
\ . \label{prtr}
\end{equation}
Here $\pen_{A}$ denotes the identity operator on the input space
$\hh_{A}$. By the cyclic property and the linearity of the trace,
formula (\ref{prtr}) implies $\Tr\bigl(\Phi(\hta)\bigr)=\Tr(\hta)$
for all $\hta\in\lnp(\hh_{A})$. To each linear map
$\Phi:{\>}\lnp(\hh_{A})\rightarrow\lnp(\hh_{B})$, one assigns its
adjoint map,
$\Phi^{\dagger}:\lnp(\hh_{B})\rightarrow\lnp(\hh_{A})$. For all
$\hta\in\lnp(\hh_{A})$ and $\htb\in\lnp(\hh_{B})$, the adjoint map
is defined by \cite{watrous1}
\begin{equation}
\bigl\langle\Phi(\hta),\htb\bigr\rangle_{\mathrm{hs}}=
\bigl\langle\hta{\,},\Phi^{\dagger}(\htb)\bigr\rangle_{\mathrm{hs}}
\ . \label{adjm}
\end{equation}
For a completely positive map (\ref{opsm}), its adjoint is
written as
$\Phi^{\dagger}(\htb)=\sum\nolimits_{n} \htk_{n}^{\dagger}\htb\htk_{n}$.
If this adjoint is trace preserving, the Kraus operators of Eq.
(\ref{opsm}) satisfy the condition
\begin{equation}
\sum\nolimits_{n}\htk_{n}\htk_{n}^{\dagger}=\pen_{B}
\ . \label{prtr1}
\end{equation}
In other words, we have $\Phi(\pen_{A})=\pen_{B}$. In this case,
the map is said to be unital \cite{bhatia07}. If a quantum map is
completely positive and the Kraus operators satisfy properties
(\ref{prtr}) and (\ref{prtr1}) the map is called bistochastic
\cite{bengtsson}, as it can be considered as an analog to the
standard bistochastic matrix, which acts in the space of
probability vectors. A quantum map $\Phi$ can be characterized
using the norm
\begin{equation}
\|\Phi\|:=\sup\bigl\{\|\Phi(\hta)\|_{\infty}:{\>}\|\hta\|_{\infty}=1\bigr\}
\ . \label{pnrm0}
\end{equation}
Let us quote here one of useful results concerning the norm of a
map. If a map $\Phi$ is positive, then
\begin{equation}
\|\Phi\|=\|\Phi(\pen)\|_{\infty}
\ , \label{pnrm1}
\end{equation}
see Bhatia \cite{bhatia07}, item 2.3.8. In terms of the
completely mixed state (\ref{cmab}), we have
$\|\Phi\|=N{\,}\|\Phi(\htr_{*})\|_{\infty}$.

The Jamio{\l}kowski isomorphism \cite{jam72} leads to another
convenient description of completely positive maps. We recall its
formulation for the symmetric case, if both dimensions are equal,
$\hh_{A}=\hh_{B}=\hh$. The principal system $A$ is extended by
an auxiliary reference system $R$ of the same dimension $N$. Let
$\{|n\rangle\}$ be an orthonormal basis in $\hh$. Making use of
this basis in both subspaces we define a maximally entangled
normalized pure state
\begin{equation}
|\phi_{+}\rangle:=\frac{1}{\sqrt{N}}{\,}\sum_{n=1}^{N}
{|n\rangle\otimes|n\rangle} \ . \label{phip}
\end{equation}
For any linear map $\Phi:{\>}\lnp(\hh)\rightarrow\lnp(\hh)$,
we assign an operator
\begin{equation}
\htvp(\Phi):=\Phi\otimes\id_{R}\bigl(|\phi_{+}\rangle\langle\phi_{+}|\bigr)
\ , \label{rdmdf}
\end{equation}
which acts on the extended space $\hh\otimes\hh$. The matrix
$\htd(\Phi)=N{\,}\htvp(\Phi)$ is usually called {\it dynamical
matrix} or {\it Choi matrix} \cite{zb04}. For any
$\htx\in{\mathcal{L}}(\hh)$, the action of the map $\Phi$ can be
recovered from $\htd(\Phi)$ by means of the relation
\cite{watrous1},
\begin{equation}
\Phi(\htx)={\Tr_{R}}{\left(\htd(\Phi){\bigl(\pen\otimes\htx^{T}\bigr)}\right)}
\ , \label{chjis}
\end{equation}
in which $\htx^{T}$ is the transpose operator to $\htx$. The
complete positivity of $\Phi$ is equivalent to the positivity of
the dynamical matrix $\htd(\Phi)$. The map $\Phi$ preserves the
trace, if and only if its dynamical matrix satisfies
\cite{watrous1}
\begin{equation}
{\Tr_{A}}{\bigl(\htd(\Phi)\bigr)}=\pen
\ . \label{trpn}
\end{equation}
In a shortened notation, we will write the dynamical matrix and
the rescaled one as $\htd_{\Phi}$ and $\htvp_{\Phi}$,
respectively. Substituting the completely mixed state
$\htr_{*}=\pen/N$ into Eq. (\ref{chjis}), we obtain
\begin{equation}
\Phi(\htr_{*})=\Tr_{R}(\htvp_{\Phi})
\ . \label{cmis}
\end{equation}
In the subsequent section we will examine a nonunitality
operator, closely related with the partial trace (\ref{cmis}).

\section{Jarzynski equality for bistochastic maps}\label{sec3}

We will consider the case, in which a thermostatted system
is operated by an external agent. It is assumed that this agent
acts according to a specified protocol. Hence, the Hamiltonian of
the system is time dependent. To formulate the Jarzynski equality,
a special kind of averaging procedure is required
\cite{tasaki99,motas10}. Initially, we describe this
procedure for arbitrary two Hermitian operators. Let us consider
operators $\hta\in\lsa(\hh_{A})$ and $\htb\in\lsa(\hh_{B})$. In
terms of the eigenvalues and the corresponding eigenstates,
spectral decompositions are expressed as
\begin{align}
\hta &=\sum\nolimits_{i} a_{i}{\,}|a_{i}\rangle\langle{a}_{i}|
\ , \label{abdc}\\
\htb &=\sum\nolimits_{j} b_{j}{\,}|b_{j}\rangle\langle{b}_{j}|
\ . \label{babdc}
\end{align}
The eigenvalues in both decompositions are assumed to be taken
according to their multiplicity. In this regard, we treat $a_{i}$
and $b_{j}$ as the labels for vectors of the orthonormal bases
$\bigl\{|a_{i}\rangle\bigr\}$ and $\bigl\{|b_{j}\rangle\bigl\}$.
Let evolution of the system in time be represented by a quantum
channel $\Phi$. If the input state is described by an eigenstate
$|a_{i}\rangle$, then the output of the channel is
$\Phi\bigl(|a_{i}\rangle\langle{a}_{i}|\bigr)$. Suppose that we
measure the observable $\htb$ in this output state. The outcome
$b_{j}$ occurs with the probability
\begin{equation}
p(b_{j}|a_{i})=\langle{b}_{j}|\Phi{\bigl(|a_{i}\rangle\langle{a}_{i}|\bigr)}|b_{j}\rangle
\ . \label{pcnji}
\end{equation}
This quantity can also be interpreted as the conditional
probability of the outcome $b_{j}$ given that the input state is
$|a_{i}\rangle$. The trace-preserving condition implies that
\begin{equation}
\sum\nolimits_{j}p(b_{j}|a_{i})={\Tr}{\Bigl(\Phi{\bigl(|a_{i}\rangle\langle{a}_{i}|\bigr)}\Bigr)}=1
\ . \label{pcnji1}
\end{equation}
The standard requirement on conditional probabilities is thus
satisfied for any quantum channel. Furthermore, we suppose that
the input density matrix $\htr_{A}$ has the form
\begin{equation}
\htr_{A}=\sum\nolimits_{i} p(a_{i})|a_{i}\rangle\langle{a}_{i}|
\ , \label{bgf}
\end{equation}
where $\sum_{i}p(a_{i})=1$. According to Bayes's rule, one
defines the joint probability distribution with elements
\begin{equation}
p(a_{i},b_{j})=p(a_{i}){\,}p(b_{j}|a_{i})
\ . \label{pabji}
\end{equation}
This is the probability that we find the system in the $i$-th
eigenstate of $\hta$ at the input and in the $j$-th eigenstate of
$\htb$ at the output. Let $f(a,b)$ be a function of two
eigenvalues. Following Ref. \cite{tasaki99},
we define the corresponding average
\begin{equation}
{\bigl\langle}{\bigl\langle}{f}(a,b){\bigr\rangle}{\bigr\rangle}:=
\sum\nolimits_{ij} p(a_{i},b_{j}){\,}f(a_{i},b_{j})
\ . \label{favm}
\end{equation}
Double angular brackets in the left-hand side denote the averaging
over the ensemble of possible pairs of measurement outcomes. A
pair of single angular brackets denotes an expectation value of an
observable $\hta$ in a state $\htr$, in consistence with the
standard notation common in quantum theory,
$\langle{\hta}\rangle={\Tr}{\bigl(\htr\hta\bigr)}$. More general forms of the
described scenario were considered in Refs.
\cite{vedral12,albash}.

The Jarzynski equality relates an averaged work with the
difference between the equilibrium free energies. Since the notion
of work pertains to a process, it cannot be represented as a
quantum observable \cite{cht11,tlh07}. A more detailed discussion
of the notion of work in the context of quantum fluctuation
theorems was recently provided by Van Vliet \cite{vvliet12}.

In any case, the energy can be measured twice, at the initial and
the final moments. The difference between outcomes of these two
measurements describes the work  performed on the system in a
particular realization \cite{tlh07}. Therefore, the averaging of
the form (\ref{favm}) is used with respect to two Hermitian
operators: the initial and the final Hamiltonians $\hth_{0}$ and
$\hth_{1}$.

Fluctuation theorems are usually obtained under the assumption
that the work is determined by projective measurements at the
beginning and the end of each run of the protocol. In several
cases one applies, however, much broader classes of quantum
measurements. Recently  Venkatesh {\it et al.} \cite{pvwt13} analyzed
fluctuation theorems for protocols in which generalized quantum
measurements are used.

In this paper we discuss the most general case of a discrete
nonunitary dynamics and consider arbitrary measurements which are
error-free in the following sense: With each outcome of a
generalized measurement, we can uniquely identify the
corresponding eigenstate of an actual Hamiltonian \cite{pvwt13}.

The system under investigation is initially prepared in the state
of the thermal equilibrium with a heat reservoir. It is convenient
to denote the inverse temperatures of the reservoir at the
beginning and at the end of the protocol, by $\beta_{0}$ and
$\beta_{1}$, respectively. In principle, these two temperatures
may differ, but in the following we will eventually discuss the
case in which both temperatures are equal. The initial density
matrix reads
\begin{equation}
\hto_{0}(\beta_{0})=Z_{0}(\beta_{0})^{-1}\exp(-\beta_{0}\hth_{0})
\ , \label{inidm}
\end{equation}
where $Z_{0}(\beta_{0})=\Tr\bigl(\exp(-\beta_{0}\hth_{0})\bigr)$
is the corresponding partition function. We further suppose that
the transformation of states of the system is represented by a
quantum channel $\Phi$, which maps the set of density matrices of
size $N$ onto itself. In general, the final density matrix
$\Phi\bigl(\hto_{0}(\beta_{0})\bigr)$ differs from the matrix
\begin{equation}
\hto_{1}(\beta_{1})=Z_{1}(\beta_{1})^{-1}\exp(-\beta_{1}\hth_{1})
\ , \label{findm}
\end{equation}
corresponding to the equilibrium at the final moment. Here, the
partition function
$Z_{1}(\beta_{1})=\Tr\bigl(\exp(-\beta_{1}\hth_{1})\bigr)$
corresponds to the state of the thermal equilibrium with the final
Hamiltonian $\hth_{1}$.

Eigenvalues of the Hamiltonians $\hth_{0}$ and $\hth_{1}$ will be
denoted by $\bigl\{\veps_{m}^{(0)}\bigr\}$ and
$\bigl\{\veps_{n}^{(1)}\bigr\}$, respectively.
Let channel $\Phi$ be unital. Using notation (\ref{favm}) for a
function of two eigenvalues, we then obtain
\begin{align}
&{\Bigl\langle}{\Bigl\langle}
{\exp}{\bigl(\beta_{0}\veps^{(0)}-\beta_{1}\veps^{(1)}\bigr)}
{\Bigr\rangle}{\Bigr\rangle}
\nonumber\\
&\equiv
\sum\nolimits_{mn}
{p}{\bigl(\veps^{(0)}_{m},\veps^{(1)}_{n}\bigr)}{\,}
{\exp}{\bigl(\beta_{0}\veps^{(0)}_{m}-\beta_{1}\veps^{(1)}_{n}\bigr)}
\nonumber\\
&=\frac{Z_{1}(\beta_{1})}{Z_{0}(\beta_{0})}
\ . \label{tasth}
\end{align}
This result was recently derived in Ref. \cite{rast13} and earlier by
Tasaki \cite{tasaki99} under a weaker assumption of a unitary
evolution. Formula (\ref{tasth}) directly leads to the Jarzynski
equality formulated for unital quantum channels.

In the approach considered the term
$W_{nm}=\veps_{n}^{(1)}-\veps_{m}^{(0)}$ is naturally identified
with the external work performed on the principal system during
the process \cite{tasaki99,deffner13}.
In the case $\beta_{0}=\beta_{1}=\beta$,
formula (\ref{tasth}) gives
\begin{equation}
{\bigl\langle}{\bigl\langle}
{\exp(-\beta{W})}
{\bigr\rangle}{\bigr\rangle}
=\exp\bigl(-\beta\Delta{F}\bigr)
\ , \label{jareq0}
\end{equation}
where the equilibrium free energies read
$F_{0,1}(\beta)=-\beta^{-1}\ln{Z}_{0,1}(\beta)$. Expression
(\ref{jareq0}) relates, on average, the nonequilibrium external
work with the difference between the equilibrium free energies,
$\Delta{F}=F_{1}-F_{0}$. Thus the above statement can be
interpreted as a version of the original Jarzynski equality
\cite{jareq97a,jareq97b}, which holds for an arbitrary unital
quantum channel.

Some other approaches to obtaining the quantum Jarzynski equality
were recently considered by Vedral \cite{vedral12} and Albash {\it et
al.} \cite{albash}. Furthermore, formula (\ref{jareq0}) was derived
in \cite{rast13} directly from Eq. (\ref{tasth}) for any
bistochastic channel. In the following we shall relax the
unitality condition and generalize this reasoning for nonunital
quantum maps.

\section{Nonunitality observable}\label{sec4}

In this section, we introduce a notion useful to analyze the
Jarzynski equality for quantum stochastic maps. In order to
characterize deviation from unitality, we are going to use the
following operator. For any trace-preserving map
$\Phi:{\>}\lnp(\hh_{A})\rightarrow\lnp(\hh_{B})$, one assigns a
traceless operator
\begin{equation}
\htg_{\Phi}:=\Phi(\htr_{*A})-\htr_{*B}
\ , \label{htxd}
\end{equation}
where $\htr_{*A}=\pen_{A}/N_{A}$ and $\htr_{*B}=\pen_{B}/N_{B}$.
This operator is Hermitian, i.e., $\htg_{\Phi}\in\lsa(\hh_{B})$,
whenever the map $\Phi$ is Hermiticity preserving. Let us derive a
useful statement about the above nonunitality operator. For given
two operators $\hta\in\lsa(\hh_{A})$, $\htb\in\lsa(\hh_{B})$ and real
parameters $\alpha$, $\beta$, we introduce the density matrices
\begin{align}
\htvr_{A}(\alpha)&:=\Tr\bigl(\exp(-\alpha\hta)\bigr)^{-1}\exp(-\alpha\hta)
\ , \label{htva}\\
\htvr_{B}(\beta)&:=\Tr\bigl(\exp(-\beta\htb)\bigr)^{-1}\exp(-\beta\htb)
\ . \label{htvb}
\end{align}
Functional forms of such a kind pertain to equilibrium in the
Gibbs canonical ensemble. We will consider average of the type
(\ref{favm}) with respect to the density matrices (\ref{htva}) and
(\ref{htvb}) at the input and output, respectively. The following
statement holds true.

\medskip

\newtheorem{lem1}{Proposition}
\begin{lem1}\label{lm1}
Let $\hta\in\lsa(\hh_{A})$, $\htb\in\lsa(\hh_{B})$, and let
$\alpha$ and $\beta$ be real numbers.  If the input state
is described by density matrix (\ref{htva}),
then the average defined in Eq.
(\ref{favm}) reads
\begin{align}
&{\Bigl\langle}{\Bigl\langle}
{\exp}{\bigl(\alpha{a}-\beta{b}\bigr)}
{\Bigr\rangle}{\Bigr\rangle}=
\nonumber\\
&\frac{N_{A}\Tr\bigl(\exp(-\beta\htb)\bigr)}{N_{B}\Tr\bigl(\exp(-\alpha\hta)\bigr)}
{\,}\left(1+N_{B}\Tr\bigl(\htvr_{B}(\beta)\htg_{\Phi}\bigr)\right)
. \label{prp1}
\end{align}
\end{lem1}

\medskip

{\bf Proof.} Using the linearity of the map $\Phi$ and Eq.
(\ref{htxd}), we obtain
\begin{align}
&\sum\nolimits_{i} p(b_{j}|a_{i})=
\langle{b}_{j}|\sum\nolimits_{i}\Phi\bigl(|a_{i}\rangle\langle{a}_{i}|\bigr)|b_{j}\rangle
\nonumber\\
&=N_{A}\langle{b}_{j}|\Phi(\htr_{*A})|b_{j}\rangle=
\frac{N_{A}}{N_{B}}+N_{A}\langle{b}_{j}|\htg_{\Phi}|b_{j}\rangle
\ . \label{sump}
\end{align}
Taking
$p(a_{i})=\Tr\bigl(\exp(-\alpha\hta)\bigr)^{-1}\exp(-\alpha{a}_{i})$
in Eq. (\ref{pabji}) and using Eq. (\ref{sump}), we represent the
left-hand side of Eq. (\ref{prp1}) in the form
\begin{align}
&\sum_{ij}\frac{\exp(-\alpha{a}_{i})}{\Tr\bigl(\exp(-\alpha\hta)\bigr)}{\>}
p(b_{j}|a_{i}){\,}\exp\bigl(\alpha{a}_{i}-\beta{b}_{j}\bigr)=
\label{ssmcp}\\
&\frac{N_{A}}{N_{B}\Tr\bigl(\exp(-\alpha\hta)\bigr)}{\>}\sum_{j}\exp(-\beta{b_{j}})
\left(1+N_{B}\langle{b}_{j}|\htg_{\Phi}|b_{j}\rangle\right)
. \nonumber
\end{align}
The latter term is easily rewritten as the right-hand side of Eq.
(\ref{prp1}). $\blacksquare$

\bigskip

If the operator $\Phi(\pen_{A})$ is proportional to $\pen_{B}$, we
have $\Phi(\pen_{A})=(N_{A}/N_{B})\pen_{B}$ by the trace
preservation. In this case, the right-hand side of Eq.
(\ref{htxd}) becomes zero. Then relation (\ref{prp1}) is reduced
to the previous result given for unital channels in Ref.
\cite{rast13}. A deviation from unitality can be quantified by
norms of the operator (\ref{htxd}). In the following, the case
$N_{A}=N_{B}=N$ will be considered. That is, the input and output
Hilbert spaces have the same dimension $N$. Note that two
different quantum channels may lead to the same nonunitality
observable. This Hermitian operator is traceless and belongs to
the space $\lsa(\hh)$ of dimensionality $N^{2}$. Due to the
Jamio{\l}kowski isomorphism, the set of quantum channels is
isomorphic with the set of their dynamical matrices satisfying Eq.
(\ref{trpn}). As the latter set has $N^4-N^2$ real dimensions
\cite{bengtsson}, there is no one-to-one correspondence between
quantum channels and the nonunitality operators. Let us estimate
the Hilbert--Schmidt norm of the operator $\htg_{\Phi}$ from
above. The following bound holds.
\smallskip

\newtheorem{lem2}[lem1]{Proposition}
\begin{lem2}\label{lm2}
Let $\hh$ be $N$-dimensional Hilbert space. If
$\Phi:{\>}\lnp(\hh)\rightarrow\lnp(\hh)$ is positive and
trace preserving, then
\begin{equation}
\bigl\|\Phi(\textup{\pen})-\textup{\pen}\bigr\|_{2}\leq\sqrt{N\bigl(\|\Phi\|-1\bigr)}
\ . \label{scnrm1}
\end{equation}
\end{lem2}

{\bf Proof.} As the map is positive and trace preserving, we have
$\Phi(\pen)\in\lsp(\hh)$ and
$\Tr\bigl(\Phi(\pen)\bigr)=\Tr(\pen)=N$. Hence, the squared
Hilbert--Schmidt norm is expressed as
\begin{equation}
\bigl\langle\Phi(\pen)-\pen{\,},\Phi(\pen)-\pen\bigr\rangle_{\mathrm{hs}}=
\|\Phi(\pen)\|_{2}^{2}-N
\ . \label{fsnrm}
\end{equation}
By positivity, we obtain
$\|\Phi(\pen)\|_{1}=\Tr\bigl(\Phi(\pen)\bigr)=N$. Lemma 3 of Ref.
\cite{rprz12} states that
$\|\hta\|_{2}^{2}\leq\|\hta\|_{\infty}{\,}\|\hta\|_{1}$ for all
$\hta\in\lnp(\hh)$. Combining these points with Eq. (\ref{fsnrm})
finally leads to
\begin{equation}
\bigl\|\Phi(\pen)-\pen\bigr\|_{2}^{2}\leq{N}\bigl(\|\Phi(\pen)\|_{\infty}-1\bigr)
{\,}. \label{scnrm}
\end{equation}
The claim (\ref{scnrm1}) follows from Eqs. (\ref{pnrm1}) and
(\ref{scnrm}). $\blacksquare$

Using Eq. (\ref{cmis}) we rewrite Eq. (\ref{scnrm1}) in the form
\begin{equation}
\bigl\|\Tr_{R}(\htvp_{\Phi})-\htr_{*}\bigr\|_{2}\leq{N}^{-1/2}\sqrt{\|\Phi\|-1}
\ . \label{bbr}
\end{equation}
In terms of the map norm, one characterizes a deviation of the
partial trace of the rescaled dynamical matrix from the completely
mixed state. If the map $\Phi$ is unital one has $\Phi(\pen)=\pen$
and $\|\Phi(\pen)\|_{\infty}=1$. Using Eq. (\ref{pnrm1}), the
right-hand side of Eq. (\ref{scnrm1}) vanishes for unital maps. On
the other hand, the condition $\|\Phi\|=1$ immediately leads to
the relation $\bigl\|\Phi(\pen)-\pen\bigr\|_{2}=0$. The latter is
equivalent to $\Phi(\pen)=\pen$, since the norm cannot be equal to
zero for a nonzero matrix, which completes the reasoning.

\medskip

\newtheorem{lem3}[lem1]{Corollary}
\begin{lem3}\label{lm3}
Let $\Phi:{\>}\lnp(\hh)\rightarrow\lnp(\hh)$ be positive and
trace preserving. If $\|\Phi\|=1$, then $\Phi$ is unital, i.e.
$\Phi(\textup{\pen})=\textup{\pen}$.
\end{lem3}

It is instructive to compare this result with the Russo--Dye
theorem. One of its formulations says that if a positive map
$\Phi$ is unital, then $\|\Phi\|=1$ (see, e.g., point 2.3.7 of Ref.
\cite{bhatia07}). In a certain sense, Corollary \ref{lm3} is a
statement in the opposite direction. Namely, if a trace-preserving
positive map $\Phi$ obeys $\|\Phi\|=1$, then it is necessarily
unital. Note that this conclusion pertains to all trace-preserving
positive maps and not only to completely positive ones. Although
legitimate quantum operations are completely positive, positive
maps without complete positivity are often used as an auxiliary
tool in the theory of quantum information. For instance, one of
the basic methods to detect quantum entanglement is formulated in
terms of entanglement witnesses and positive maps \cite{HHH96}.

Due to (\ref{scnrm1}), a deviation from unitality is
characterized by the difference between the norm
$\|\Phi\|$ and unity.
 It is possible to find an upper bound for this
quantity in terms of the dimension $N$ of the Hilbert space.
From  Eq. (\ref{npqr}), we obtain
$\|\Phi(\pen)\|_{\infty}\leq{N}=\|\Phi(\pen)\|_{1}$. Combining
this with Eqs. (\ref{pnrm1}) and (\ref{scnrm1}) we get
\begin{equation}
\bigl\|\Phi(\pen)-\pen\bigr\|_{2}\leq \sqrt{N(N-1)}
\ .
\label{cnrmn}
\end{equation}
This inequality is valid for all trace-preserving positive maps
$\Phi:{\>}\lnp(\hh)\rightarrow\lnp(\hh)$, including quantum
channels with the same input and output spaces. Rescaling the
above bound by the dimensionality, we have
\begin{equation}
\|\htg_{\Phi}\|_{2}\leq\left(\|\Phi(\htr_{*})\|_{\infty}-N^{-1}\right)^{1/2}
\leq \sqrt{1-1/N}
\ . \label{rscmn}
\end{equation}
Thus, the Hilbert--Schmidt norm of the operator
$\htg_{\Phi}=\Phi(\htr_{*})-\htr_{*}$ is strictly less than $1$.
The left-hand side of Eq. (\ref{rscmn}) can be interpreted as the
Hilbert--Schmidt distance between $\Phi(\htr_{*})$ and $\htr_{*}$.
This distance is maximal if  inequality (\ref{rscmn}) is saturated
so the output state  $\Phi(\htr_{*})=|\psi\rangle\langle\psi|$ is
pure. This is the case for the map generated by Kraus operators
$\htk_{n}=|\psi\rangle\langle{n}|$, where the
$\bigl\{|n\rangle\bigr\}$ is an orthonormal basis in $\hh$. Such a
map represents a complete contraction to a pure state: For any
state $\htr$ one has $\Phi(\htr)=|\psi\rangle\langle\psi|$. Taking
$|\psi\rangle$ as a ground state one can describe in this way the
process of spontaneous emission in atomic physics.

Systems near the thermal equilibrium can be treated as ergodic in
the following sense: Any quantum state can be reached, directly or
indirectly, from any other state. In this regard, the completely
contracting channel has opposite properties, as for any initial
state only a single state  $|\psi\rangle$ can be reached during
the process.

Using representation (\ref{blrx}), the nonunitality operator
$\htg_{\Phi}$ can be represented in terms of its generalized Bloch
vector $\bau$ \cite{KR01,zb04} with components
$\tau_{j}=\Tr\bigl(\htg_{\Phi}\htl_{j}\bigr)$. Therefore we arrive
at a handy expression for the nonunitality operator,
$\htg_{\Phi}=(1/2){\,}\bau\cdot\hat{\blm}$, where $\hat{\blm}$
denotes the $(N^{2}-1)$-dimensional vector of generators of
${\mathrm{SU}}(N)$.
We also obtain an upper bound for the modulus of the
Bloch vector,
\begin{equation}
|\bau|\leq\sqrt{2}\left(\|\Phi(\htr_{*})\|_{\infty}-N^{-1}\right)^{1/2}
\leq \sqrt{2-2/N}
\ . \label{rscmnt}
\end{equation}
It follows from Eq. (\ref{rscmn}) and the expression for the
squared Hilbert--Schmidt norm
$\langle\htg_{\Phi}{\,},\htg_{\Phi}\rangle_{\mathrm{hs}}=
(1/2)\sum\nolimits_{j=1}^{N^{2}-1}\tau_{j}^{2}$. In the case
$N=2$, the bound (\ref{rscmnt}) gives $|\bau|\leq 1$ for all
quantum channels, as in the normalization used in this work the
set of one-qubit states forms the Bloch ball of radius one.
The right-hand side of Eq. (\ref{rscmnt}) tends to $\sqrt{2}$ for
large $N$.

\section{Jarzynski equality for arbitrary stochastic maps}\label{sec5}

We now apply Eq. (\ref{prp1}) in a physical set-up corresponding
to the context of the Jarzynski equality \cite{jareq11}. One
assumes that a thermostatted system is acted upon by an
external agent, which operates according to the prescribed
protocol. The principal system is assumed to be prepared initially
in the state of the thermal equilibrium with a heat reservoir.
 As before we denote the initial and the final inverse temperatures
of the reservoir by $\beta_{0}$ and $\beta_{1}$.
 Therefore, the input state is described by density matrix (\ref{inidm}).
According to the actual process, the final density matrix
$\Phi\bigl(\hto_{0}(\beta_{0})\bigr)$ may differ from the state
(\ref{findm}), which corresponds to the equilibrium at the final
moment. Considering the same system, we also assume that both
dimensions are equal, $N_{A}=N_{B}=N$. By substitutions, relation
(\ref{prp1}) leads to the equality
\begin{align}
&{\Bigl\langle}{\Bigl\langle}
{\exp}{\bigl(\beta_{0}\veps^{(0)}-\beta_{1}\veps^{(1)}\bigr)}
{\Bigr\rangle}{\Bigr\rangle}=
\nonumber\\
&\frac{Z_{1}(\beta_{1})}{Z_{0}(\beta_{0})}
\left(1+N\Tr\bigl(\hto_{1}(\beta_{1})\htg_{\Phi}\bigr)\right)
. \label{tasthg}
\end{align}
For unital quantum channels, the term
$N\Tr\bigl(\hto_{1}(\beta_{1})\htg_{\Phi}\bigr)$ is zero,
 so formula (\ref{tasthg}) forms an extension of the previous result
(\ref{tasth}) and for a unitary evolution it reduces to the result
of Tasaki \cite{tasaki99}. The right-hand side of Eq.
(\ref{tasthg}) depends not only on equilibrium properties of the
system but also on the realized process. The authors of Ref.
\cite{talhag09} emphasized such a feature in connection with
nonequilibrium relations for the exponentiated internal energy
and heat.

Note also that concrete details of the realized quantum process
are represented by means of a single operator $\htg_{\Phi}$. To
formulate fluctuation relation (\ref{tasthg}), no additional
characterization of the map is required. In this regard, we need
not to specify a kind of coupling between the principal system and
its environment.

The quantity $W_{nm}=\veps_{n}^{(1)}-\veps_{m}^{(0)}$ can be
identified with the external work performed on the principal
system during a process \cite{deffner13,tasaki99}. For the case
$\beta_{0}=\beta_{1}=\beta$, formula (\ref{tasthg}) leads to a
generalized form of the Jarzynski equality for arbitrary,
nonunital quantum channels,
\begin{equation}
{\bigl\langle}{\bigl\langle}
{\exp(-\beta{W})}
{\bigr\rangle}{\bigr\rangle}
={\exp}{\bigl(-\beta\Delta{F}\bigr)}
{\left(1+N\Tr\bigl(\hto_{1}(\beta)\htg_{\Phi}\bigr)\right)}
{\,}. \label{jareq0g}
\end{equation}
This is the central result of the present work. The correction
term $N\Tr\bigl(\hto_{1}(\beta)\htg_{\Phi}\bigr)$ characterizes a
deviation induced by the nonunitality of a map. In general, this
term can be positive, equal to zero, or negative. If the
channel $\Phi$ is unital, then the operator $\htg_{\Phi}$ and the
correction term are equal to zero, so the standard form
(\ref{jareq0}) of the Jarzynski equality is recovered. It is
essential to note that the correction term may vanish also for
nonunital quantum channels, provided
$\Tr\bigl(\hto_{1}\htg_{\Phi}\bigr)=0$.

For a convex function $y\mapsto\exp(-\beta{y})$, the Jensen
inequality implies
${\bigl\langle}{\bigl\langle}{\exp(-\beta{W})}{\bigr\rangle}{\bigr\rangle}
\geq{\exp}{\bigl(-\beta\langle\langle{W}\rangle\rangle\bigr)}$.
Combining this with Eq. (\ref{jareq0g}) gives
\begin{equation}
\langle\langle{W}\rangle\rangle
\geq\Delta{F}-\beta^{-1}{\,}{\ln}{\left(1+N\Tr\bigl(\hto_{1}(\beta)\htg_{\Phi}\bigr)\right)}
{\,}. \label{lbww}
\end{equation}
This inequality provides a lower bound on the average work
performed on a driven quantum system. If the correction term is
strictly negative, the right-hand side of Eq. (\ref{lbww}) is
strictly larger than $\Delta{F}$. The latter bound is commonly
known and takes place for quasistatic processes. On the other
hand, positivity of the correction term will reduce this bound. It
is an evidence for the fact that the averaged external work may,
in principle, be less than $\Delta{F}$, provided the macroscopic
process investigated is sufficiently far from unitality.

It is instructive to discuss limiting cases of high and low
temperatures. For sufficiently high temperatures, if
$|\beta\veps_{n}|\ll1$ with some typical value $\veps_{n}$, the
correction term can be expanded as
\begin{equation}
N\Tr\bigl(\hto_{1}(\beta)\htg_{\Phi}\bigr)=
N{Z}_{1}^{-1}(\beta){\left(-\beta\Tr\bigl(\hth_{1}\htg_{\Phi}\bigr)+O(\beta^{2})\right)}
{\,}. \label{hhtm}
\end{equation}
Since the nonunitality operator $\htg_{\Phi}$ is traceless, the
expansion starts with the first-order term with respect to
$\beta$. If $\Tr\bigl(\hth_{1}\htg_{\Phi}\bigr)=0$, the right-hand
side of Eq. (\ref{hhtm}) also vanishes in the first order. Within
this approximation, the standard form (\ref{jareq0}) and its
consequences remain valid. For very low temperatures, the
correction term can be expressed in terms of the ground-state
energy, $\veps_{0}=\min\bigl\{\veps_{n}^{(1)}\bigr\}$. If this
state is nondegenerate, we approximately write
\begin{equation}
N\Tr\bigl(\hto_{1}(\beta)\htg_{\Phi}\bigr)=N\langle\veps_{0}|\htg_{\Phi}|\veps_{0}\rangle
\ . \label{lltm}
\end{equation}
Neglected terms are of the order of
$O\bigl(\exp(-\beta\Delta\veps)\bigr)$, where
$\beta\Delta\veps\gg1$ and $\Delta\veps>0$ is a typical distance
between nearest-neighbor levels, for instance, the energy
difference between the ground state and the first excited state.
Up to a high accuracy the deviation from the unitality is
represented by a single matrix element
$\langle\veps_{0}|\htg_{\Phi}|\veps_{0}\rangle$, as probabilities
of excited states becomes negligible for low temperatures. In
general, this matrix element characterizes a difference of the
matrix element $\langle\veps_{0}|\Phi(\rho_{*})|\veps_{0}\rangle$
from the equiprobable value $1/N$. If the ground state is not
involved in the undergoing process, the correction term vanishes.
Thus, in the low-temperature limit the standard form (\ref{jareq0})
may be adequate, even if the process itself is generally far from
equilibrium. We also observe that the right-hand side of Eq.
(\ref{lltm}) does not depends on the temperature.

The above results can be put into the context of the heat transfer
between two quantum systems. The composite Hilbert space
$\hh_{AB}=\hh_{A}\otimes\hh_{B}$ is a tensor product of the
Hilbert spaces $\hh_{A}$ and $\hh_{B}$ of individual systems. Let
us rewrite Eq. (\ref{prp1}) so the initial state of the
composite system reads
\begin{align}
\htvr_{AB}&:=\Tr\bigl(\exp(-\alpha\hta)\bigr)^{-1}{\,}\Tr\bigl(\exp(-\beta\htb)\bigr)^{-1}
\nonumber\\
& \times\exp(-\alpha\hta)\otimes\exp(-\beta\htb)
\ , \label{abvr}
\end{align}
where $\hta\in\lsa(\hh_{A})$ and $\htb\in\lsa(\hh_{B})$. Using
an observable
$\htc:=\alpha\hta\otimes\pen_{B}+\pen_{A}\otimes\beta\htb$,
we may rewrite operator (\ref{abvr}) as
\begin{equation}
\htvr_{AB}=\Tr\bigl(\exp(-\htc)\bigr)^{-1}{\,}\exp(-\htc)
\ . \label{abvr1}
\end{equation}

Assume now  that the evolution of the composite system is
represented by a quantum channel
$\Psi:{\>}\lnp(\hh_{AB})\rightarrow\lnp(\hh_{AB})$. By
corresponding substitutions in Eq. (\ref{prp1}), we obtain
\begin{equation}
{\Bigl\langle}{\Bigl\langle}
{\exp}{\bigl(\alpha{a}+\beta{b}-c\bigr)}
{\Bigr\rangle}{\Bigr\rangle}
=1+N_{A}N_{B}\Tr\bigl(\htvr_{AB}\htg_{\Psi}\bigr)
{\,} , \label{abvr2}
\end{equation}
where $\htg_{\Psi}=\Psi(\htr_{*AB})-\htr_{*AB}$. The operator
$\htg_{\Psi}$ vanishes for unital channels and the right-hand side
of Eq. (\ref{abvr2}) becomes equal to unity as discussed in Ref.
\cite{rast13}. We now consider the following situation. Two
separated systems are initially prepared in equilibrium with the
inverse temperatures $\beta_{0}$ and $\beta_{1}$, respectively.
Then the combined system is initially described by the tensor
product
$\htom_{01}:=\hto_{0}(\beta_{0})\otimes\hto_{1}(\beta_{1})$.
Making use of Eq. (\ref{abvr2}) we obtain
\begin{align}
&{\Bigl\langle}{\Bigl\langle}
{\exp}{\bigl(\beta_{0}(\veps^{(0)}-\veps^{\prime(0)})+\beta_{1}(\veps^{(1)}-\veps^{\prime(1)})\bigr)}
{\Bigr\rangle}{\Bigr\rangle}
\nonumber\\
&=1+N_{A}N_{B}\Tr\bigl(\htom_{01}\htg_{\Psi}\bigr)
{\,}. \label{abvr3}
\end{align}
Following Ref. \cite{tasaki99} we introduce the quantity
\begin{equation}
\Delta{S}:=
{\Bigl\langle}{\Bigl\langle}
{\beta_{0}(\veps^{\prime(0)}-\veps^{(0)})+\beta_{1}(\veps^{\prime(1)}-\veps^{(1)})}
{\Bigr\rangle}{\Bigr\rangle}
\ . \label{abvs}
\end{equation}
As the terms
${\bigl\langle}{\bigl\langle}{\veps^{\prime(0)}-\veps^{(0)}}{\bigr\rangle}{\bigr\rangle}$ and
${\bigl\langle}{\bigl\langle}{\veps^{\prime(1)}-\veps^{(1)}}{\bigr\rangle}{\bigr\rangle}$ are
average variations of self-energies of the two subsystem,
quantity (\ref{abvs}) describes a contribution of these variations
into a change of the total entropy.
 Combining Eq. (\ref{abvr3}) with the Jensen inequality finally gives a bound
$\Delta{S}\geq{-\ln}{\left(1+N_{A}N_{B}\Tr\bigl(\htom_{01}\htg_{\Psi}\bigr)\right)}$.
If variations of the inverse temperatures are sufficiently small
and the contributions of interaction energy are negligible then
the quantity  $\Delta{S}$ provides an estimate of changes of the
total entropy of the system \cite{tasaki99}. If the correction
term is strictly negative, then $\Delta{S}>0$. Negativity of the
correction term also implies
$\langle\langle{W}\rangle\rangle>\Delta{F}$. Since contributions
on the interaction energy are small enough, a perturbative
description of the process is reasonable. On the other hand,
positivity of the correction term implies $\Delta{S}<0$. In such a
case, contributions of the interaction can be relevant, so
quantity (\ref{abvs}) does not provide  a legitimate estimate for
changes of the total entropy.

\section{Exemplary nonunital quantum maps}\label{sec6}

In this section we discuss some simple nonunital quantum channels
and analyze the correction term present in the Jarzynski equality
(\ref{jareq0g}).  Analyzed channels describe the effects of the
energy loss from an interacting quantum system and can be
considered as a generalization of the amplitude damping channel
\cite{nielsen,bengtsson}.

\subsection{Two--level system}

\begin{figure*}
\includegraphics[height=7.8cm,angle=0]{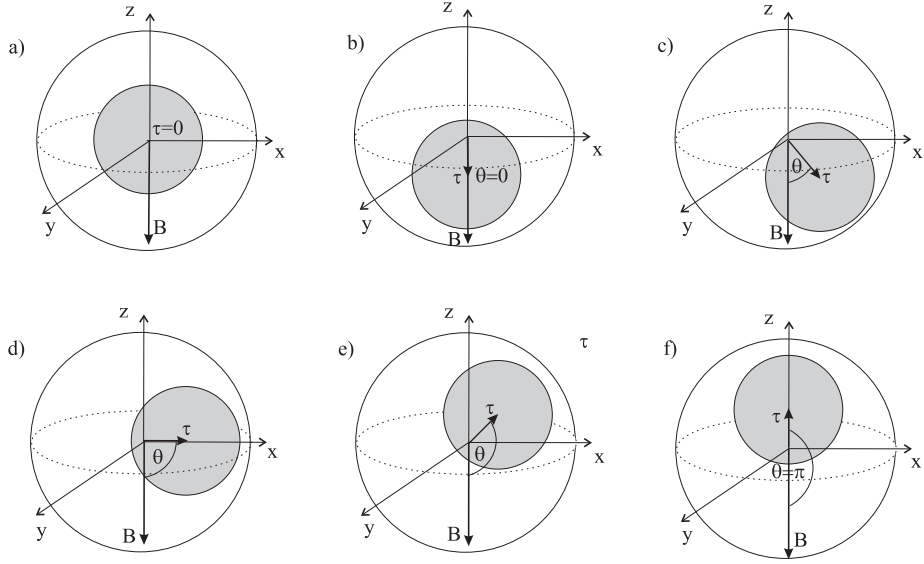}
\caption{\label{fig1} One--qubit quantum channels acting on the
Bloch ball. Correction term in the Jarzynski equality
(\ref{jareq0g}) depends on the product $\bau\cdot\mib$ and
vanishes in the unital case (a) and in the case (d), for which both
vectors are perpendicular. Correction term is maximal in the cases
(b) and (f), for which these vectors are parallel.}
\end{figure*}

Consider the simplest case $N=2$ representing a one qubit system.
Let a magnetic moment with spin $1/2$ and a charge $-e$ be in
contact with a thermal bath at the inverse temperature $\beta$.
The corresponding Hamiltonian reads
\begin{equation}
\hth_{1}=-\mu_{B}{\,}\mib\cdot\hat{\bsg}
\ . \label{hml2}
\end{equation}
Here, $\mu_{B}={e}\hbar/(2mc)$ is the Bohr magneton, $\mib$ is the
vector of an external field, and $\hat{\bsg}$ is the vector of the
three Pauli matrices. Suppose that the time evolution of the
system is represented by the amplitude damping channel
\cite{nielsen,bengtsson} described by the Kraus operators
\begin{equation}
\htk_{0}=
\begin{pmatrix}
\sqrt{1-p}{\,} & 0 \\
0 & 1
\end{pmatrix}
{\,}, \qquad
\htk_{1}=
\begin{pmatrix}
0 & 0 \\
\sqrt{p} & 0
\end{pmatrix}
{\,}, \label{kk12}
\end{equation}
with $p\in[0,1]$. For this channel the image of the maximally
mixed state, $\Phi(\htr_{*})$, can be represented by the Bloch
vector $\bau=(0,0,-p)$. The length $|\bau|=p$ of the Bloch vector
characterizes the degree of the nonunitality of the map. In the
original map $\Phi$ described by the operators $\htk_{0}$ and
$\htk_{1}$ the translation vector $\bau$ is parallel to the axis
$z$, but this can be changed, if the nonunitary dynamics is
followed by an arbitrary unitary rotation,
\begin{equation}
\htr\mapsto\htr^{\prime\prime}=\htu\Phi(\htr)\htu^{\dagger}
\ , \label{hruu}
\end{equation}
where $\htu\in{\mathrm{U}}(2)$. Then the rotated translation vector $\bau$
can take an arbitrary orientation with respect to the magnetic
field, pointing along the $z$ axis; see Fig. \ref{fig1}. In
general, the nonunitality observable (\ref{htxd}) reads therefore
$\htg_{\Phi}=(1/2){\,}\bau\cdot\hat{\bsg}$. The correction term in
Eq. (\ref{jareq0g}) becomes then
\begin{equation}
2{\,}\Tr\bigl(\hto_{1}(\beta)\htg_{\Phi}\bigr) =
\Tr\bigl(\hto_{1}(\beta)\bau\cdot\hat{\bsg}\bigr)=
p\tanh(\beta\mu_{B}B)\cos\theta
\ , \label{cort2}
\end{equation}
where $B=|\mib|$ and $\theta$ denotes the angle between the Bloch
vector $\bau$ and the magnetic field $\mib$. Another
interpretation of the angle $\theta$ follows from the scalar
product in the Hilbert--Schmidt space of operators,
\begin{equation}
\langle\hth_{1}{\,},\htg_{\Phi}\rangle_{\mathrm{hs}}=
-p{\,}\mu_{B}B\cos\theta
\ . \label{cst}
\end{equation}
In this example, the product $|p\cos\theta|$ is a natural measure
of deviation from unitality. Writing the correction term in a
coordinate-independent manner, we obtain the Jarzynski equality
\begin{align}
&{\Bigl\langle}{\Bigl\langle}
{\exp}{\bigl(-\beta{W}\bigr)}
{\Bigr\rangle}{\Bigr\rangle}=
\nonumber\\
&{\exp}{\bigl(-\beta\Delta{F}\bigr)}
\Bigl(1+\tanh(\beta\mu_{B}B)B^{-1}\bau\cdot\mib\Bigr)
{\>}. \label{jareq2g0}
\end{align}

For high temperatures, the right-hand side of (\ref{jareq2g0})
reads
${\exp}{\bigl(-\beta\Delta{F}\bigr)}\Bigl(1+\beta\mu_{B}\bau\cdot\mib\Bigr)$
due to $\tanh(\beta\mu_{B}B)\approx\beta\mu_{B}B$. If the
translation vector is perpendicular to the external field,
$\bau\perp\mib$, the correction term vanishes for arbitrary values
of $\beta$. This case provides a concrete physical example, in
which the standard form of Jarzynski equality holds for nonunital
quantum channels. The absolute value of the correction term is
maximal for $\bau\parallel\mib$. Some configurations of the
vectors $\bau$ and $\mib$ are shown in Figs. \ref{fig1}(a)--\ref{fig1}(f).
A size of the correction term also depends on the temperatures and
it is small in the high-temperature limit. In the low-temperature
limit, $\tanh(\beta\mu_{B}B)\to1$ and the right-hand side of Eq.
(\ref{cort2}) is merely reduced to $p\cos\theta$. When the
parameters $\beta$, $B$, and $\theta$ are fixed, the correction
term becomes maximal for $p=1$. Here we deal with a spontaneous
emission channel, which maps all inputs to some prescribed pure
state. In the sense of Eq. (\ref{hruu}), this prescribed state can
be chosen arbitrarily.

With the above example, we can return to the discussion of the
notion of work in the context of quantum fluctuation relations
\cite{cht11,vvliet12,tlh07}. In the nonunital cases in Figs.
\ref{fig1}(b) and \ref{fig1}(f), one has an energy shift of the ``center of mass'' of
the set of states, resulting in maximization of the correction
term. Such an effect means that one takes into account, on
average, the work against the magnetic field. In Fig. \ref{fig1}(d), the
shift is orthogonal to the field, whence no energy change occurs
and the correction term vanishes.

Overall, the average energy cost due to the work against the field
depends on both the length of $\bau$ and its direction, as it is
described by the last, nonunitality term in Eq. (\ref{jareq0g}).
Note that this term, vanishing for instance for any unitary
evolution, is not related to average changes of the von Neumann
entropy of the quantum state during nonunitary processes. As the
evolution $\Phi$ is not unitary, pure states may be converted into
mixed states, or mixed states into pure, due to the interaction
with an environment. Thus, to produce a nonunitary map $\Phi$,
some work has to be exchanged between the principal system and the
environment.

\subsection{Three--level system}

We now consider a generalized amplitude damping channel acting on
a $N=3$ quantum state and parametrized by two real numbers,
$p,q\in[0,1]$. The map is described by a set of three Kraus
operators. It contains a single diagonal matrix,
$\htk_{0}={\mathrm{diag}}\bigl(\sqrt{1-p},\sqrt{1-q},1\bigr)$ and
two nondiagonal matrices,
\begin{equation}
\htk_{1}=
\begin{pmatrix}
0 & 0 & 0 \\
0 & 0 & 0 \\
\sqrt{p} & 0 & 0
\end{pmatrix}
{\,}, \quad
\htk_{2}=
\begin{pmatrix}
0 & 0 & 0 \\
0 & 0 & 0 \\
0 & \sqrt{q} & 0
\end{pmatrix}
{\,}. \label{kk123}
\end{equation}
For this channel we find the nonunitality observable
$\htg_{\Phi}=(1/3){\,}{\mathrm{diag}}\bigl(-p,-q,p+q\bigr)$. As
the choice $p=q=0$ leads to the identity map, we assume in the
following that $p\neq0$ or $q\neq0$. For a massive particle with
spin $1$ and charge $e$, we write the final Hamiltonian,
\begin{equation}
\hth_{1}=\frac{e}{mc}{\>}\mib\cdot\hat{\blj}
\ . \label{hml3}
\end{equation}
Let us take the axis $z$ such that the component $\hat{J}_{z}$
commutes with $\htg_{\Phi}$, i.e.,
$\hat{J}_{z}=\hbar{\>}{\mathrm{diag}}\bigl(+1,0,-1\bigr)$. In
their common eigenbasis, the two other components of the spin are
expressed as
\begin{equation}
\hat{J}_{x}=\frac{\hbar}{\sqrt{2}}
\begin{pmatrix}
0 & 1 &  0 \\
1 & 0 & 1 \\
0 & 1 & 0
\end{pmatrix}
{\,}, \quad
\hat{J}_{y}=\frac{\hbar}{\sqrt{2}}
\begin{pmatrix}
0 & -i &  0 \\
i & 0 & -i \\
0 & i & 0
\end{pmatrix}
{\,}. \label{jxy}
\end{equation}
In this case, an expression for the correction term is more
complicated than Eq. (\ref{cort2}). For sufficiently
high temperatures, when $\beta$ is small, one has a simple
approximation,
\begin{equation}
3{\,}\Tr{\bigl(\hto_{1}(\beta)\htg_{\Phi}\bigr)}=
(2p+q)\beta{\,}\frac{e\hbar}{3mc}{\>}{\mathbf{e}_{z}}\cdot\mib+O(\beta^{2})
\ , \label{cort3}
\end{equation}
where ${\mathbf{e}}_{z}$ is the unit vector of the axis $z$. If
$\mib\perp{\mathbf{e}}_{z}$, the correction term vanishes in the
first order with respect to $\beta$. In this respect, expression
(\ref{cort3}) is analogous to Eq. (\ref{cort2}). On the other
hand, for a generic value of the temperature, the correction term
typically differs from zero. When we fix $\beta$ and $\mib$, then
the correction term in the first order becomes maximal for
$p=q=1$. As in the above case $N=2$, this choice gives a complete
contraction to some pure state. In the low-temperature limit, we
can rewrite Eq. (\ref{cort3}) in the form
\begin{equation}
3{\,}\Tr{\bigl(\hto_{1}(\beta)\htg_{\Phi}\bigr)}
={\left(p+\frac{q}{2}\right)}\cos\theta+\frac{q}{8}{\,}{\bigl(1+3\cos2\theta\bigr)}
\ , \label{cort3l}
\end{equation}
where $\theta$ is the angle between ${\mathbf{e}}_{z}$ and
$\mib$. As mentioned above, the right-hand side of Eq.
(\ref{cort3l}) neglects contributions of order
$\exp\bigl(-\beta{e}\hbar{B}/(mc)\bigr)$ with very large values of
the exponent $\beta{e}\hbar{B}/(mc)$.

\subsection{$N$--level system}

Consider the following process defined for an arbitrary 
$N$-dimensional space. Let $I$ and $J$ be two sets of indices such
that $I\cap{J}=\emptyset$ and
$I\cup{J}=\bigl\{1,2,\ldots,N\bigr\}$. The map is described by a
set of the Kraus operators. The first of them is chosen to be
diagonal in the eigenbasis of the Hamiltonian,
\begin{equation}
\htk_{0}:=\sum\nolimits_{m\in{I}} z_{m}|\veps_{m}\rangle\langle\veps_{m}|
+\sum\nolimits_{n\in{J}} |\veps_{n}\rangle\langle\veps_{n}|
\ . \label{k0df}
\end{equation}
For given $n\in{I}$ and arbitrary $m\neq{n}$,
we define further operators,
\begin{equation}
\htk_{mn}:=a_{mn}|\veps_{m}\rangle\langle\veps_{n}|
\ , \label{kmndf}
\end{equation}
for which
$\htk_{mn}^{\dagger}\htk_{mn}=|a_{mn}|^{2}{\,}|\veps_{n}\rangle\langle\veps_{n}|$
and
$\htk_{mn}\htk_{mn}^{\dagger}=|a_{mn}|^{2}{\,}|\veps_{m}\rangle\langle\veps_{m}|$.
For all $n\in{I}$ we impose a restriction,
\begin{equation}
|z_{n}|^{2}+\sum\nolimits_{m\neq{n}}|a_{mn}|^{2}=1
\ , \label{imrs}
\end{equation}
whence $|z_{n}|\leq1$ and $|a_{mn}|\leq1$. Hence, condition
(\ref{prtr}) is satisfied, i.e., the considered quantum operation
is trace preserving. For brevity, we put positive numbers
$y_{m}=\sum_{n\neq{m}}|a_{mn}|^{2}$ for each
$m\in\bigl\{1,2,\ldots,N\bigr\}$. An explicit form of all Kraus
operators allows us to find the image of the identity operator,
\begin{align}
\Phi(\pen)&=\sum\nolimits_{m\in{I}} \left(|z_{m}|^2+y_{m}\right)|\veps_{m}\rangle\langle\veps_{m}|
\nonumber\\
&+\sum\nolimits_{n\in{J}}(1+y_{n}){\,}|\veps_{n}\rangle\langle\veps_{n}|
\ . \label{phpn}
\end{align}
Therefore the nonunitality observable $\htg_{\Phi}$ is diagonal
with elements $x_{m}=N^{-1}\bigl(|z_{m}|^2+y_{m}-1\bigr)$ for
$m\in{I}$ and elements $x_{n}=N^{-1}y_{n}$ for $n\in{J}$. The
correction term in Eq. (\ref{jareq0g}) can be then written as
\begin{align}
N\Tr{\bigl(\hto_{1}(\beta)\htg_{\Phi}\bigr)}&=N\left(\sum\nolimits_{n}\exp(-\beta\veps_{n})\right)^{-1}
\nonumber\\
&\times\sum\nolimits_{n}x_{n}\exp(-\beta\veps_{n})
\ . \label{xxnn}
\end{align}

Note that the right-hand side of Eq. (\ref{xxnn}) represents the
correction term for arbitrary $\htg_{\Phi}$. In this case, we
merely replace $x_{n}$ with the diagonal matrix element
$\langle\veps_{n}|\htg_{\Phi}|\veps_{n}\rangle$ with respect to
the Hamiltonian eigenbasis.  The correction term is not
uniquely defined by a given quantum channel. Hence, effects of
nonunitality in the Jarzynski equality and related fluctuation
relations in some cases  may be modeled by a generalized amplitude
damping channel in the described form. It is possible, provided
the diagonal element
$\langle\veps_{n}|\htg_{\Phi}|\veps_{n}\rangle$ of the operator
$\htg_{\Phi}$ can be represented in terms of the above introduced
numbers $z_{n}$ and $y_{n}$.

The above two damping channels acting on $N=2$ and $N=3$ systems
are particular cases of the general scheme. For instance, matrices
(\ref{kk12}) are obtained for $I=\{1\}$ and $J=\{2\}$ with
$z_{1}=\sqrt{1-p}$, $a_{21}=\sqrt{p}$. Matrices (\ref{kk123}) and
diagonal
$\htk_{0}={\mathrm{diag}}\bigl(\sqrt{1-p},\sqrt{1-q},1\bigr)$ are
recovered by setting $I=\{1,2\}$ and $J=\{3\}$ with
$z_{1}=\sqrt{1-p}$, $z_{2}=\sqrt{1-q}$, $a_{31}=\sqrt{p}$,
$a_{32}=\sqrt{q}$. Another version of the damping channel for a
three-level system is described for $I=\{1\}$ and $J=\{2,3\}$.
Taking $z_{1}=\sqrt{1-p}$, $a_{21}=\sqrt{q}$, $a_{31}=\sqrt{p-q}$,
we obtain the Kraus matrices
$\htk_{0}={\mathrm{diag}}\bigl(\sqrt{1-p},1,1\bigr)$,
\begin{equation}
\htk_{21}=
\begin{pmatrix}
0 & 0 & 0 \\
\sqrt{q} & 0 & 0 \\
0 & 0 & 0
\end{pmatrix}
{\,}, \quad
\htk_{31}=
\begin{pmatrix}
0 & 0 & 0 \\
0 & 0 & 0 \\
\sqrt{p-q}{\,} & 0 & 0
\end{pmatrix}
{\,}. \label{kk1233}
\end{equation}
These operators lead to the diagonal matrix
$\Phi(\pen)-\pen={\mathrm{diag}}\bigl(-p,q,p-q\bigr)$. Therefore
operators (\ref{kk1233}) allow a controlled shift of the
population between the levels of the system.

\section{Concluding remarks}

In this work we formulated Jarzynski equality (\ref{jareq0g}) for
a quantum system described by an arbitrary stochastic map. This is
a direct generalization of earlier results obtained for unital
maps \cite{albash,rast13}, for which the maximally mixed state is
preserved. We derived a correction term which compensates the
nonunitality of the map and attempted to estimate its relative
size. Furthermore, it was shown that the correction term vanishes
if the nonunitality observable is perpendicular, in the sense of the
Hilbert--Schmidt scalar product, to the Hamiltonian of the system.
Hence, expression (\ref{jareq0}) obtained previously remains valid
also for certain cases of nonunital maps provided the
nonunitality does not influence the average energy of the system.

The results are exemplified on a simple model of the damping
channel. For a two-level system, the correction term depends on
the nonunitality measured by the length of the translation vector
$\bau$ and its orientation with respect to the vector of magnetic
field. The latter determines the Hamiltonian of the system. When
other parameters are fixed, the translation vector is the longest
in the case of complete contraction to a pure state. For the
considered two-level example, such a map leads to the maximal
relative size of the correction term. However, if the translation
vector is perpendicular to the field, the correction term vanishes
irrespectively of the length of the translation vector.

As a by-product of our study we introduced the nonunitality
operator $G_{\Phi}$  associated with a given quantum operation
$\Phi$  and analyzed its properties. Some useful bounds for its
norm have been established. Furthermore, we presented a broad
class of nonunitary dynamics acting in the set of quantum states
of an arbitrary finite dimension $N$, which can serve as a
generalization of the one-qubit amplitude damping channel.

\bigskip

The authors are very grateful to Peter H\"{a}nggi for a fruitful
correspondence, several valuable comments, and access to his
unpublished notes. We are thankful to Daniel Terno for useful
discussions. Financial support by the Polish National Science
Centre, Grant No. DEC-2011/02/A/ST1/00119 (K.\.Z.), is gratefully
acknowledged.

\end{document}